\input harvmac
\Title{\vbox{\baselineskip12pt\hbox{EFI-96-04}
\hbox{hep-th/9602049}}}
{New Principles for String/Membrane Unification}

\centerline{David Kutasov and Emil Martinec\footnote{*}{Supported 
in part by Dept. of Energy grant DEFG02-90ER-40560.  } 
}
\vskip .5cm
\centerline{\it Enrico Fermi Inst. and Dept. of Physics}
\centerline{\it University of Chicago}
\centerline{\it 5640 S. Ellis Ave., Chicago, IL 60637}
\vskip 2cm
 
\noindent
The target space theory of the N=(2,1) heterotic string 
may be interpreted as a theory of gravity coupled to matter
in either $1+1$ or $2+1$ dimensions.  
Among the target space theories in $1+1$ dimensions are 
the bosonic, type II, and heterotic string world sheet 
field theories in a physical gauge.  
The $2+1$ dimensional version describes
a consistent quantum theory of 
supermembranes in $10+1$ dimensions.  
The unifying framework for all of these vacua is a theory of $2+2$
dimensional self-dual geometries embedded in $10+2$ dimensions.
There are also indications that the N=(2,1) string
describes the strong coupling dynamics of compactifications 
of critical string theories to two dimensions,
and may lead to insights about the fundamental degrees of
freedom of the theory.

\Date{2/96}

\def\journal#1&#2(#3){\unskip, \sl #1\ \bf #2 \rm(19#3) }
\def\andjournal#1&#2(#3){\sl #1~\bf #2 \rm (19#3) }

\def\ie{{\it i.e.}}
\def\eg{{\it e.g.}}

\def\etal{{\it et.al.}}

\def\sst{\scriptscriptstyle}

\def\frac#1#2{{#1\over#2}}
\def\coeff#1#2{{\textstyle{#1\over #2}}}
\def\half{\frac12}
\def\hf{{\textstyle\half}}
\def\ket#1{|#1\rangle}

\def\d{\partial}

\def\inbar{\,\vrule height1.5ex width.4pt depth0pt}
\def\IC{\relax\hbox{$\inbar\kern-.3em{\rm C}$}}
\def\IR{\relax{\rm I\kern-.18em R}}
\def\IP{\relax{\rm I\kern-.18em P}}
\def\Z{{\bf Z}}

%
%
\def\np#1#2#3{Nucl. Phys. {\bf B#1} (#2) #3}
\def\pl#1#2#3{Phys. Lett. {\bf #1B} (#2) #3}
\def\prl#1#2#3{Phys. Rev. Lett. {\bf #1} (#2) #3}
\def\physrev#1#2#3{Phys. Rev. {\bf D#1} (#2) #3}

\def\cqg#1#2#3{Class. Quant. Grav. {\bf #1} (#2) #3}

\catcode`\@=11
\def\slash#1{\mathord{\mathpalette\c@ncel{#1}}}
\overfullrule=0pt

\def\FF{{\cal F}}

\def\JJ{{\cal J}}

\def\PP{{\cal P}}

\def\VV{{\cal V}}

\def\XX{{\cal X}}

\def\lam{\lambda}

\def\underrel#1\over#2{\mathrel{\mathop{\kern\z@#1}\limits_{#2}}}

\catcode`\@=12


%

\def\ket#1{\left| #1\right\rangle}

\def\det{{\rm det}}
\def\tr{{\rm tr}}

\def\det{{\rm det}}
\def\exp{{\rm exp}}

\def\twoone{{(2,1)}}
\def\twozero{{(2,0)}}

\def\tpt{{2+2}}

\def\xbar{{\bar x}}
\def\zbar{{\bar z}}
\def\tbar{{\bar\theta}}
\def\psibar{{\bar\psi}}
\def\phibar{{\bar\phi}}

\def\ibar{{\bar i}}
\def\jbar{{\bar j}}
\def\kbar{{\bar k}}

\def\dbar{{\bar \d}}
\def\Dbar{{\bar D}}
\def\Xbar{{\bar X}}


\newsec{Introduction}

Recent progress in string duality (for a recent status report,
see \ref\hull{C. Hull, hep-th/9512181.})
indicates that there is but one unified theory whose low-energy
manifestations include all string theories, as well as
eleven-dimensional supergravity.  This poses a puzzle,
since the latter seems to be a theory of quantum membranes
(sometimes called M-theory), whereas the usual world-volume
analysis of such theories 
\ref\membrev{E. Bergshoeff, E. Sezgin and P. Townsend,
Ann. Phys. {\bf 185} (1988) 330.}\
as 2+1 (super-) gravity coupled to matter does not lead to a 
renormalizable quantum theory.  Moreover, the classical equations
of motion are highly nonlinear and difficult to solve, hardly 
a promising starting point.  Trying to turn these shortcomings
into virtues, some have speculated that, since the eleven-dimensional
theory has no small dimensionless coupling, there ought not to 
be a sensible theory of `free' quantum membranes.
However, deeper understanding ought to arise if we can find
new principles by which to construct the quantum theory of
membranes, hopefully in tandem with a new approach to string theory
that incorporates all of its various manifestations.  We believe that
N=\twoone\ heterotic string theory shows us the way to such
new principles.

In this paper, we are going to revive an old idea of Green's
\ref\green{M. Green, \np{293}{1987}{593}.}:
That string theories whose target space dynamics is effectively
two-dimensional have a low-energy effective action which is
the worldsheet field theory of another string.
This is all but guaranteed by the fact that the 
target space of string theory always contains gravity, and
two-dimensional gravity coupled to matter is by definition
string theory.  
What is not guaranteed is that one obtains
critical string theory in this way; in fact, most of Green's models
were noncritical strings.  
As one increases the worldsheet gauge symmetry of string theory,
one comes closer to realizing critical strings in the
two-dimensional target space.  It turns out that N=\twoone\
local supersymmetry is the magical principle which generates all
critical string theories, as well as membranes.
For lower supersymmetry, one generally has (a) infinite towers of
massive states, leading to complicated effective interactions
for the massless modes; and (b) the wrong number of massless fields.
N=(2,2) supersymmetry seems too restrictive.

N=2 strings have been studied by many authors over the years
(for a review, see
\ref\marcus{N. Marcus,
talk at the Rome String Theory Workshop (1992); hep-th/9211059.};
for early work, see
\nref\neqtwostr{M. Ademollo \etal, \pl{62}{1976}{105}, 
\np{111}{1976}{77};
E. S. Fradkin and A. A. Tseytlin, \pl{106}{1981}{63}, \pl{162}{1985}{295};
A. D'Adda and F. Lizzi, \pl {191}{1987}{85};
S. Mathur and S. Mukhi, \physrev{36}{1987}{465}; \np{302}{1988}{130}.}%
\refs{\neqtwostr,\green}).
The N=2 string properly lives in two 
complex target dimensions, with 4+0 or 2+2 signature.
Ooguri and Vafa 
\nref\ovone{H. Ooguri and C. Vafa, Mod. Phys. Lett. {\bf A5} (1990) 1389.}%
\nref\ovtwo{H. Ooguri and C. Vafa, \np{361}{1991}{469}.}%
\nref\ovthree{H. Ooguri and C. Vafa, \np{367}{1991}{83}.}%
\refs{\ovone-\ovthree}\
systematically investigated the geometry for these theories.
This geometry is self-dual, admitting a twistorial formulation
\nref\sokatchev{E. Witten, hep-th/9410052;
A. Galperin, E. Sokatchev, hep-th/9504124.}%
\refs{\ovtwo,\sokatchev}.
Thus the N=(2,2) closed and open string theories describe
some sort of quantization of self-dual gravity and self-dual
Yang-Mills, respectively.  
The heterotic N=\twoone\ and (2,0)
theories add a new wrinkle.  The right-moving
worldsheet gauge invariance requires 2+2 signature, which would
spoil the no-ghost theorem for the left-moving tower
of modes.  Ooguri and Vafa therefore add two target dimensions to
the left-movers, such that the left movers are in 10+2 or 26+2
dimensions; and simultaneously enlarge the left-moving gauge
principle to include a null current algebra, to factor out
the unwanted additional timelike modes.  Any null current will do;
if we choose its orientation entirely within the subspace
common to both left- and right-movers, the effective target
space dynamics is reduced from 2+2 to 1+1 dimensions.
Thus heterotic N=2 strings actually do describe the 1+1 dimensional
target space kinematics employed in \green.  In addition, 
Ooguri and Vafa point out another possibility: One may
orient the null current partly in the 2+2 spacetime and
partly in the `internal space' of the left-movers.
This effectively removes only one (timelike) direction
from the 2+2 target, leaving a 2+1 dimensional target space
dynamics.  This theory contains a kind of gravity inherited
from self-dual gravity in four dimensions, together with
matter fields depending on the details of the construction.

The version of the \twoone\ heterotic
string with 1+1 dimensional target has several vacua, depending
on a choice of GSO projection.  
Below we will exhibit examples having (a) 24 bosons; (b)
8 bosons, 8 left-moving, and 8 right-moving fermions with (8,8)
supersymmetry; and (c) 8 bosons, 8 right-moving fermions, 
and 16 chiral left-moving bosons with (8,0) supersymmetry.
These target space theories are {\it free} and have no massive
excitations; they are, respectively, 
the bosonic, type II, and heterotic string world sheet 
field theories in a physical gauge.  
The construction of the target space heterotic string involves
an orbifold that can only be seen from the $2+2$ dimensional
point of view.

The \twoone\ heterotic string with $2+1$ dimensional 
target space has a version containing
8 bosons, 8 fermions, having N=8 supersymmetry; and
provides a promising candidate for a consistent quantum theory of 
supermembranes in $11$ dimensions.  
Two features of this theory stand out: First, the classical
theory is integrable, being a restricted form of 2+2
self-dual gravity coupled to self-dual matter; second, the
\twoone\ string gives us some prescription for the quantum 
theory (it is not ruled out that one might be able to quantize
directly the target space effective field theory).

Our plan is as follows: In section 2, we 
briefly review the N=\twoone\
heterotic string and establish our conventions.  In section 3
we consider models whose dynamics appears 1+1 dimensional in the 
2+2 target space by choosing the null current orthogonal to the
left-moving internal space.  We exhibit choices of GSO projection
which yield bosonic, type II, and heterotic strings as the effective
target space field theories.  In section 4, we turn to the
models which appear 2+1 dimensional, with null current partly in
the internal space.  A particular choice of GSO projection gives
a candidate for the membrane of eleven-dimensional supergravity;
another gives a bosonic membrane in 27 dimensions.
All discussion in these two sections
will take place in the context of flat target geometries.
In section 5, we briefly examine the sigma model describing the 
curved space generalization of the theory.
We conclude in section 6 with a list of comments and speculations.


\newsec{N=(2,1) Heterotic Strings}

In this paper we will be mostly discussing $(2,1)$ strings,
heterotic strings with $N=2$ local supersymmetry for the
right movers, and $N=1$ supersymmetry
for the left movers. Since many aspects
of the construction are common to all the models we will
discuss, we will start with a review of some of the
relevant features of the right-moving $N=2$ and the left-moving
$N=1$ chiral sectors. This will also serve to establish our
notation. In the next sections we will discuss different
ways of putting the two sectors
together. We will be using rather heavily 
the technology developed in \ref\fms{D. Friedan, E. Martinec and
S. Shenker, \np{271}{1986}{93}.}; the reader may want to consult
this reference for further details. For more details on
$N=2$ strings, see e.g. \refs{\neqtwostr-\ovthree}.

\subsec{The right movers: $N=2$ strings.}

In flat spacetime $\IR^{2,2}$,
the right movers of the N=\twoone\ heterotic
string are four real scalar fields $x^\mu$,
$\mu=0,1,2,3$ with signature $(-,-,+,+)$, and 
their superpartners under the $N=2$ superconformal
algebra, $\bar\psi^\mu$. We will sometimes bosonize
the fermions $\bar\psi$, as:
\eqn\psiboson{
  \eqalign{\psibar^0\pm\psibar^3=&e^{\pm i\bar H_1}\cr
	\psibar^1\pm\psibar^2=&e^{\pm i\bar H_2}\ ,\cr}
}
and will freely switch between the fermionic 
and bosonic representations.
The $N=2$ superconformal generators can be written as:
\eqn\rightNeqtwo{\eqalign{
  \bar T = & -\hf\dbar x\dbar x - \psibar\dbar\psibar\cr
  \bar G^\pm =& e^{\pm i\bar H_1}\bar\partial(x^0\pm x^3)+
 e^{\pm i\bar H_2}\bar\partial(x^1\pm x^2)\cr
    \bar J =& i\bar\partial \bar H_1+i\bar\partial \bar H_2 ,\cr
}}

To construct physical states in the covariant BRST
formalism one also needs the ghosts for the
gauge algebra generated by  $\bar T$, ${\bar G}_{\pm}$, $\bar J$;
these will be denoted by $(\bar b,\bar c)$, 
$({\bar \beta}_\pm, {\bar \gamma}_\pm)$,
and $(\bar{\tilde b},\bar{\tilde c})$, respectively. In particular
vertex operators will turn out to depend on the fields
$\bar\phi_\pm$ arising from the bosonization of $\bar\beta_\pm$,
$\bar\gamma_\pm$ in the usual way \fms: 
${\bar\beta}_\pm{\bar\gamma}_\pm=\dbar\bar\phi_\pm$. 

Most (but not all) vertex operators we will need will have the same
form as far as the right-moving $N=2$ string part is concerned:
\eqn\vvtwo{\eqalign{
V_{-1}(k)=&e^{-\bar\phi_+-\bar\phi_-}e^{ik\cdot x}\cr
V_0(k)=& \bar G^+_{-\half}\bar G^-_{-\half}e^{ik\cdot x}\cr}}
with the first line describing the vertex operator in the
$-1$ picture, and the second, in the $0$ picture.

Three point functions of physical states reduce to products
of right- and left-moving parts; the $N=2$ right-moving
part contributes
\eqn\threept{\langle V_{-1}(k)V_{-1}(p)V_0(r)\rangle=(k_0+k_3)(p_0-p_3)
+(k_1+k_2)(p_1-p_2)-(k\leftrightarrow p).}
We will use below the fact that when one reduces all the momenta
such that $k_0+k_3=p_0+p_3=r_0+r_3=0$, the amplitude \threept\
vanishes.

\subsec{The left-moving $N=1$ sector.}

To describe the left-moving sector we need to find a $\hat c=10$ 
$N=1$ superconformal field theory (SCFT). The $2+2$ dimensional spacetime
coordinates $x^\mu$ (which are non-compact and hence shared by the
left and right movers), and their superpartners, $\psi^\mu$ form
a $\hat c=4$ SCFT. Naively we need an additional $\hat c=6$, but there
is a subtlety \ovthree. 
As explained in the introduction, ghost elimination requires us
to gauge a null $U(1)$ 
supercurrent on the left side. The ghosts corresponding to this
$U(1)$ carry $\hat c=-2$ so overall we are looking for an internal
$\hat c=8$ ($c=12$) SCFT. A convenient representation 
of a large class of such theories is provided by twenty-four free
fermions, $\lambda^a$ \ref\pierce{D. Pierce, hep-th/9601125.}. 
The total $N=1$ superconformal algebra for 
the left movers is:
\eqn\leftNeqone{\eqalign{
  T = &-\hf\d x\d x - \psi\d\psi-\lambda\d\lambda\cr
  G = &\psi\d x + \coeff16 f_{abc}\lambda^a\lambda^b\lambda^c\ .\cr
}}
Here $f_{abc}$ are the structure 
constants of some semisimple group of dimension 24,
under which the $\lambda^a$
transform as the adjoint representation. 
As for the $N=2$ case, there are ghosts $b,c$, $\beta, \gamma$ 
needed to gauge \leftNeqone; as well as ghosts of the supersymmetric
$U(1)$ current algebra, which we will suppress. Some vertex operators
depend on the bosonized $\beta, \gamma$ ghost $\beta\gamma=\partial\phi$.
A special case of \leftNeqone\
that we will use is obtained by bosonizing sixteen of the twenty-four
fermions, and describing the internal SCFT in terms of eight
left-moving scalars $y^a$
(living on the $E_8$ torus for modular invariance)
and eight left-moving fermions, $\lambda^a$.

As mentioned above, to define a vacuum of the $N=(2,1)$ string one 
needs to identify in the left-moving $N=1$ SCFT a null $U(1)$
supersymmetric current algebra, that is then gauged.
There are two fundamentally inequivalent choices:

\item{1)} If the $U(1)$ lies entirely in the `spacetime', 
$\IR^{2,2}$ part of the theory, one finds a theory that 
effectively is dimensionally reduced from $2+2$ to $1+1$ dimensions.
We will use the supercurrent 
$(\psi^0+\psi^3)+\theta(\partial x^0+\partial
x^3)$, which reduces the target space to $\IR^{1,1}$ parametrized
by $(x^1, x^2)$.

\item{2)} If the $U(1)$ lies partly in spacetime and partly
in the internal theory, $J=\partial x^0+J_{\rm internal}$,
we find a theory in $1+2$ non-compact dimensions, with a second
compact time dimension $x^0$.

\noindent In addition to the choice above,
the theories are also parametrized
by the choice of GSO projection on the fermions $\lambda^a$ etc.
In the next sections we will describe the physics of the different
theories that arise from various natural choices. 


\newsec{Target Space Strings}

Above we saw that $N=(2,1)$ theories fall into two broad
classes, depending on the choice of the null $U(1)$ 
(actually $\IR$) super-current algebra 
that is gauged on the left-moving, $N=1$ side.
If the $U(1)$ lies purely in the four dimensional ``spacetime''
part of the $(2,1)$ sigma model, we find theories whose target
space is 1+1 dimensional. These theories
will be discussed in this section. A $U(1)$ that lies partly 
in spacetime and partly in the internal supeconformal field theory
leads to theories in 2+1 dimensional target space; such theories will 
be described in the next section.
Eventually we will see that, regardless of such a choice of $U(1)$
embedding, the theory is 
more properly thought of  in \tpt\ dimensions.

String theory is a theory of spacetime gravity. Therefore, when 
the target space of the $(2,1)$ string is two-dimensional, we expect
the spacetime dynamics to describe matter coupled to two-dimensional
gravity. We will in fact see that the two-dimensional gravity systems
that arise correspond to various models of critical strings, with the
two-dimensional target space of the $(2,1)$ heterotic string serving
as the world sheet of another string. This by itself is not too 
surprising -- it is well known that the dynamics on the world sheet 
of a string describes two-dimensional (super-) gravity, and conversely,
essentially any $2d$ gravity theory can be thought of as a world 
sheet string theory.
What is nevertheless interesting here is that:

\item{a)} The $2d$ gravity systems that arise in the target
space of the $N=2$ heterotic string seem to correspond to 
{\it critical} string theories.

\item{b)} Critical (super-) string theories with different
gauge principles can be obtained in this way, starting from
closely related $(2,1)$ heterotic string vacua.
In particular, small variations in the GSO projection 
on the world sheet of the $N=2$ string
give rise in target space to the world sheet theories of twenty-six
dimensional bosonic, ten dimensional type II and heterotic strings.
These describe different (super-) gravities on the world sheet --
the target space of the $N=2$ string --
and until recently were considered quite different.

\noindent
In what follows we describe the constructions that lead to the different 
types of strings in target space. We use the notation of section 2.

\subsec{Bosonic strings.}

Perhaps the simplest theory of $(2,1)$ strings is one in which 
the spin structures of all $24$ left-moving fermions $\{\lambda^a\}$
are identified, and one projects onto states with $(-)^{F_L}=1$.
It is easy to write down the (world sheet) 
torus partition sum \pierce:
\eqn\partt{
  Z(\tau)={1\over2}\left[
	\left({\theta_3\over\eta}\right)^{12}-
	\left({\theta_4\over\eta}\right)^{12}-
	\left({\theta_2\over\eta}\right)^{12}
	\right].
}
$Z(\tau)$ is modular invariant, holomorphic in the
standard fundamental domain of the modular group, $\FF$, hence it is 
constant: $Z(\tau)=24$. Indeed, the spectrum of physical states
consists in this case of $24$ massless scalars arising from the Neveu-
Schwarz (NS) sector of the internal superconformal field theory. 
These scalars live on the 
Lorentzian target space $\IR^{1,1}$ parametrized by $x^1$, $x^2$
(recall that we are gauging the null current $J=\partial( x^0+x^3)$).
We will denote $(x^1\pm x^2)$ by $x^\pm$ for brevity. 

The vertex operators of the scalars, in the $-1$ picture for all
three $\beta, \gamma$ systems in the problem, are:
\eqn\scver{
  V^a=e^{-\phi}\lambda^a e^{-\bar\phi_+-\bar\phi_-}
	e^{ik\cdot x}\ ;\qquad k_+k_-=0,\quad a=1,\cdots,24.
}
Since the ground state energy of the Ramond (R) sector for the left 
movers $\lambda^a$ is positive, no physical states arise from that
sector. Note that we have set the momentum in the $x^0+x^3$ direction to
zero. We could have multiplied $V^a$ \scver\ by $\exp[iq(x^0+x^3)]$
but since $q$ is a null vector, this would have no effect on the
dynamics, leading to an equivalent representation of $V^a$. 
The gauged $U(1)$ 
symmetry allows one to identify the NS and R sectors
of the right-moving $N=2$ superconformal field theory; 
when shifting R $\rightarrow$ NS one has to shift 
the momentum in the $(x^0+x^3)$ direction, but that 
is inconsequential as argued above.

The scattering amplitudes of the $24$ scalars $V^a$ on the sphere
vanish on shell, because of special properties of $N=2$ strings --
see \threept\ and the subsequent discussion.
Hence, the target space action is free:
\eqn\starg{
  S_{\rm target}={1\over\lambda^2}\int d^2x\partial_+V^a \partial_-V^a.
}
Here $\lambda$ is the string coupling constant of the
`fundamental' $N=(2,1)$ string. Unlike more complicated 
string theories, where actions like \starg\ are only valid
at low energy, \starg\ is exact.

The action $S_{\rm target}$ describes free fields, but we expect
the theory to contain target space gravity as well. 
The metric can be parametrized, after fixing diffeomorphisms, by its
conformal factor, $g_{\alpha\beta}=\eta_{\alpha\beta}e^\rho$,\foot{A
different parametrization of the metric, adapted to the
Kahler geometry of target space, will be explored in section 5.} 
and in addition we expect a dilaton $\sigma$ related to the string 
coupling by $\langle e^\sigma\rangle=\lambda$. The fields $\sigma$ and
$\rho$ do not describe propagating degrees of freedom -- they can be
gauged away. Eq. \starg\ describes the theory in a physical gauge; the 
covariant action (in target space) is:
\eqn\scov{
  S_{\rm target}=\int d^2x e^{-2\sigma}\sqrt{g}\left[
	 g^{\alpha\beta}\partial_\alpha V^a
	\partial_\beta V^a-\coeff14
	g^{\alpha\beta}\partial_\alpha\sigma
	\partial_\beta\sigma+R+\Lambda\right]
} 
It is natural to suggest that one should think of \scov\ as
the world sheet action for a $26$ dimensional bosonic string with 
$\{V^a\}$ the transverse dimensions of the string, $\sigma, \rho$
the longitudinal ones. In the target space of the $(2,1)$ string,
$x^\pm$ is reinterpreted as the world sheet of that bosonic string. 
The (2,1) string field space is reinterpreted as $26$ dimensional 
spacetime.  The bosonic string has infinite extent, 
filling the $(\rho, \sigma)$ plane.

Note that the bosonic string whose world sheet is the target space
of the $(2,1)$ string has string tension $T_B\simeq 1/\lambda^2$
(see \scov). Thus, string perturbation theory in the $(2,1)$ theory
gives the ``sigma model'' ($\alpha^\prime$) expansion for the
bosonic string. 
One can ask what is the string coupling $\lambda_B$ of this bosonic
string. {}From \scov\ it appears that the bosonic string dilaton 
$\Phi$ (not to be confused with the $N=(2,1)$ string dilaton, $\sigma$)
satisfies $\Phi\simeq 1/\lambda^2$. Hence a natural guess is \green
\eqn\stcop{
  \lambda_B\equiv e^\Phi\simeq a e^{b/\lambda^2}
}
with some positive constants $a,b$. When the $N=2$  string is weakly 
coupled, the target space bosonic string is strongly coupled.

Further insight into the structure of the theory can be obtained 
by studying it on a target space two-torus\foot{One can either
consider a Lorentzian torus or study the theory with positive
signature.}. A two-dimensional torus has two complex moduli; $T$
which parametrizes the complex structure, and $U$ which measures the
area and $B$ field, measured in units of $\alpha'$. 
The partition sum of the $(2,1)$ string on 
a target space torus with moduli $T$, $U$ is given (perturbatively)
by:
\eqn\ztt{
  \log Z_{\rm target}(T,U,\lambda)=\sum_{h=1}^\infty
	\lambda^{2h-2}Z_h(T,U),
}
where $Z_h$ is the world sheet genus $h$ path integral with target
space a torus with moduli $T$ and $U$. In addition one expects
non-perturbative corrections in $\lambda$, which we will discuss later.
The leading term in the expansion \ztt, corresponding to a world sheet
torus, is obtained from the non-compact partition sum \partt\ by replacing 
a factor of $1/\tau_2$ by a sum over momenta and windings on the 2-torus:
\eqn\zone{
  Z_1(T,U)={1\over2}\cdot 24\cdot\int_\FF {d^2\tau\over\tau_2^2 }
	\sum_{(p_L, p_R)}q^{{1\over2}p_L^2}
	\bar q^{{1\over2}p_R^2}
}
with $q=\exp[2\pi i\tau]$; and ($T_2\equiv Im(T)$; $U_2\equiv Im(U)$):
\eqn\mmm{\eqalign{
  p_L=&{1\over\sqrt{2T_2 U_2}}\left(n_1+n_2T+(m_1+m_2T)U\right)\cr
  p_R=&{1\over\sqrt{2T_2 U_2}}\left(n_1+n_2T+(m_1+m_2T)\bar U\right)\cr
}}
This integral has been computed in 
\ref\dkl{L. Dixon, V. Kaplunovsky, and J. Louis, \np{355}{1991}{649};
see also A. Font, L. Ibanez, D. L\"ust, and F. Quevedo,
\pl{245}{1990}{401}; S. Ferrara, N. Magnoli, T. Taylor,
and G. Veneziano, \pl{245}{1990}{409}.}
with the result (ignoring a $T, U$ independent infrared
divergent constant):
\eqn\zzone{
  Z_1=-24\log\left(\sqrt{T_2}|\eta(T)|^2
	\sqrt{U_2}|\eta(U)|^2\right).
}
Comparing to \ztt\ we conclude that the leading term in 
the target torus partition sum is:
\eqn\resz{
  Z_{\rm target}(T, U, \lambda)=
	\left({1\over\sqrt{T_2}|\eta(T)|^2}\right)^{24}
	\left({1\over\sqrt{U_2}|\eta(U)|^2}\right)^{24}
	+O(\lambda^2).
}
This looks like the partition sum of two decoupled sets
of $24$ scalars living on 2-tori with moduli $T$, $U$ respectively.
These scalars are the $\{V^a\}$.  A second set of scalars appears
due to the existence of winding modes.  On-shell vertex operators
take the form
\eqn\momwind{\eqalign{
  V_{++}^a=&\Xi^a e^{i(k_+x^+ + \kbar_+\xbar^+)}\cr
  V_{--}^a=&\Xi^a e^{i(k_-x^- + \kbar_-\xbar^-)}\cr
  V_{+-}^a=&\Xi^a e^{i(k_+x^+ + \kbar_-\xbar^-)}\cr
  V_{-+}^a=&\Xi^a e^{i(k_-x^- + \kbar_+\xbar^+)}\ ,\cr
}}
where $\Xi^a=e^{-\phi}\lambda^a e^{-\bar\phi_+-\bar\phi_-}$.
Naively, the momentum operators $V_{++}$, $V_{--}$
only couple to deformations $T$
of the torus complex structure,
\eqn\tdef{
  T=\d x^+\dbar x^+\qquad,\qquad \bar T=\d x^-\dbar x^-\ ,
}
while winding operators $V_{+-}$, $V_{-+}$
only couple to the scale and theta parameter
\eqn\udef{
  U=\d x^+\dbar x^-\qquad,\qquad \bar U=\d x^-\dbar x^+\ .
}
The momentum and winding vertices $V^a$ correspond to, respectively, 
chiral-chiral and chiral-antichiral
operators of the global N=(2,2) superconformal algebra allowed
by a flat target space.
Their decoupling is probably a consequence
of properties of c-c and c-a rings in two-dimensional $N=2$ 
superconformal field theory.  We leave a more detailed
analysis of this issue to future work.

It is interesting to think about the structure of possible corrections to
\resz\ of higher order in $\lambda$ \ztt. Suppose first that all such 
corrections vanished. That would imply that the scalars $\{V^a\}$
are non-compact, and the bosonic target space string lives in 
$26$ non-compact dimensions. It is difficult to decide apriori
from the $N=(2,1)$ string point of view whether the scalars $V^a$
are compact or not. We will later argue that there is (weak) 
evidence that they in fact live on a $24$ -- torus, 
in which case the corrections to 
\resz\ should turn it into:
\eqn\newzt{
  Z_{\rm target}(T, U, \lambda)=Z_T(T, \lambda) Z_U(U, \lambda),
}
with 
\eqn\zttnew{
  Z_T(T,\lambda)={1\over|\eta(T)|^{48}}\sum_{
	(\PP_R,\PP_L)\in\Gamma_{24,24}}
	(e^{2\pi i T})^{{1\over2}\PP_R^2}
	(e^{-2\pi i \bar T})^{{1\over2}\PP_L^2}
}
where $\PP_{R,L}$ are target-space momenta in the even self-dual
lattice $\Gamma_{24,24}$, such that $Z_T$
is modular invariant in $T$. 
A similar structure should hold for $U$.

Since all length scales in string theory are measured in units
of $\alpha^\prime$, and we saw that for the target space bosonic string
$\alpha^\prime_B\simeq \lambda^2$, at weak $N=2$ coupling $\lambda\to0$
the size of the $24$--torus diverges. Thus, the fact that 
\resz\ appears to describe non-compact scalars $V^a$ may be an artifact 
of the weak coupling limit $\lambda\to0$. The precise determination of 
the lattice $\Gamma_{24,24}$ requires the knowledge of higher orders in 
the perturbative expansion \ztt.

Knowledge of the perturbative series \ztt\ allows one to 
calculate the non-perturbative corrections by using \ztt, \newzt\ and 
modular invariance of $Z_{\rm target}$ in $T$, $U$ separately. These
corrections, that are due to winding modes around the $24$ dimensional 
torus (solitons in the original $N=2$ string language), should go like
$\exp(-c/\lambda^2)$.

One can also ask about the higher genus
amplitudes of the target space string theory; one might imagine
they can be analyzed by formulating the N=(2,1) string on
a target space of the form $T^*\Sigma$, the cotangent space
of a Riemann surface \ovtwo, or a suitable \twoone\ sigma-model
generalization with null Killing vector.

\subsec{Type II strings.}

To construct a $(2,1)$ string theory whose target 
space dynamics describes a type II world sheet
we modify slightly the GSO projection for the fermions
$\{\lambda_a\}$. We split them into a set of eight
and a set of sixteen. Performing separate GSO projections on
the two sets we find a theory 
that can be alternatively described in terms of $8$ scalars
$y^a$ compactified on the $E_8$ torus, and 
$8$ fermions $\lambda_a$ ($a=1,\cdots,8$).
The left-moving $N=1$ superconformal generator is
\eqn\spc{
  G=\psi_\mu \d x^\mu + \lambda^a\partial y^a.
}
This theory has $N=2$ superconformal symmetry and, one may hope,
spacetime supersymmetry. This is indeed the case as is clear from 
the world sheet torus partition sum:
\eqn\prt{Z(\tau)={1\over4}
\left[
\left({\theta_3\over\eta}\right)^4-
\left({\theta_4\over\eta}\right)^4-
\left({\theta_2\over\eta}\right)^4
\right]
\left[
\left({\theta_3\over\eta}\right)^8+
\left({\theta_4\over\eta}\right)^8+
\left({\theta_2\over\eta}\right)^8
\right].}
The two factors in \prt\ correspond to the contributions
of $\lambda^a$, $y^a$, respectively.
The first factor in \prt\ vanishes due to a well known 
identity; the theory has the same number of bosons and
fermions. It is easy to construct the spacetime supercharges.
Denoting by $H$ the scalar obtained by bosonizing $\psi^\pm$, the
superpartners of $x^\pm$ under the left-moving superconformal
algebra, and by 
$S_\alpha$, $S_{\bar\alpha}$ the dimension $1/2$ vertex
operators creating the $s$, $c$ spinors of the $SO(8)$ 
associated with $\lambda_a$, they are:
\eqn\schg{
\eqalign{
	Q_\alpha=&\oint dz e^{-{\phi\over2}+{i\over2}H}S_\alpha\cr
	Q_{\bar\alpha}=&\oint dz e^{-{\phi\over2}-{i\over2}H}
	S_{\bar\alpha}\cr
}}
The $16$ supercharges \schg\ form an $(8,8)$ supersymmetry algebra
in target space:
\eqn\susyal{
\eqalign{
	\{Q_\alpha, Q_\beta\}=&\delta_{\alpha\beta}P^+\cr
	\{Q_{\bar\alpha}, Q_{\bar\beta}\}=&
	\delta_{\bar\alpha\bar\beta}P^-\cr
	\{Q_\alpha, Q_{\bar\beta}\}=&0\cr
}}
The spectrum consists of eight massless scalars, with vertex operators:
\eqn\scalrs{
	V^a=e^{-\phi}\lambda^a e^{-\bar\phi_+-\bar\phi_-}
	e^{ik\cdot x};\;\;k_+k_-=0,\;a=1,\cdots,8.
}
and eight right and left-moving fermions, with vertex operators:
\eqn\frms{
\eqalign{
	\chi_\alpha=&e^{-{\phi\over2}+{i\over2}H}
	S_\alpha e^{-\bar\phi_+-\bar\phi_-}
	e^{ik\cdot x};\;\;k^+=0\cr
	\bar\chi_{\bar\alpha}=&e^{-{\phi\over2}-{i\over2}H}
	S_{\bar\alpha} e^{-\bar\phi_+-\bar\phi_-}
	e^{ik\cdot x};\;\;k^-=0\cr
}}
The matter content and symmetry structure is consistent 
with an interpretation of the target space theory as describing
a type II world sheet in a physical gauge, with $V^a$ the 
eight transverse dimensions, and $\chi_\alpha$, $\bar\chi_{\bar\alpha}$
the static gauge Green-Schwarz fermions. 
Because one is in a physical gauge, half of the spacetime supersymmetries
are realized nonlinearly.
As in the bosonic construction,
one can ask whether the $\{V^a\}$ are compact. From the form of the vertex
operators for $V^a$ \scalrs\ it seems natural to guess that they are 
compactified on the $E_8$ torus. That would also mean that there
is no distinction between IIA and IIB strings (the two are related
by $T$ duality).

Many aspects of the discussion are similar to the bosonic case analyzed
in the last subsection, and will not be repeated. In particular, the
theory is free in physical gauge (just like the Green -- Schwarz
superstring), and one can determine the
string coupling and $\alpha^\prime$ of the type II string that is
obtained in this construction in terms of the underlying $(2,1)$ string
parameters.

\nref\kmo{D. Kutasov, E. Martinec, and M. O'Loughlin, hep-th/9603116.}%
\subsec{Heterotic strings\foot{\rm For an improved treatment and
further discussion, see \kmo.}.}

In the discussion so far, the dynamics of the theory in the
$x^0$ and $x^3$ directions was taken to be trivial.
However, one should actually think of these theories as intrinsically
$2+2$ dimensional.  This point is driven home by the construction
of heterotic strings in our framework, to which we now turn.
Ho\v rava and Witten \ref\horwit{P. Ho\v rava and E. Witten, 
hep-th/9510209.}
found that the $E_8\times E_8$ heterotic string could be
described as an orbifold of M-theory compactified on $S^1/\Z_2$.
This leads us to try to describe the heterotic string as
a $\Z_2$ orbifold of the (2,1) string.  Since we are in static
gauge, the M-theory world-volume is identified with three of the
eleven spacetime coordinates.  Therefore an orbifold of the (2,1)
theory should correspond to the above spacetime orbifold.

This time, we split the $24$ 
fermions into three sets of $8$, denoted by 
$\{\lambda^a_i\}$ $a=1,\cdots,8$, $i=1,\cdots,3$.
The contribution of these fields to the left-moving supercurrent 
\leftNeqone\ is
\eqn\fermG{
  G=\sum_a \lambda^a_1\lambda^a_2\lambda^a_3\ .
}
corresponding to \leftNeqone\ with group $[SU(2)]^8$.
We want to preserve only half of the supersymmetries \schg\ (constructed
as before from the spin fields of the first group of eight fermions,
$\{\lambda_1\}$). The unbroken 
supercharges, say $Q_\alpha$,
form an $(8,0)$ supersymmetry 
algebra on $\IR^{1,1}$ (see \susyal).
The $\Z_2$ twist we employ is left-right asymmetric on the
(2,1) worldsheet, leaving the right-movers untouched, and
twisting the left-moving part of the $\IR^{2,2}$ coordinates
$(x^0,x^3)\rightarrow -(x^0,x^3)$;
this twist is coupled with a GSO
projection on the three sets of fermions $\lambda^a_i$.
This GSO projection acts as $(-1)^F$ separately on each
of the three sets of left-moving fermions.  
Denote the twist field\foot{Because $(x^0,x^3)$ 
are non-compact, there is a unique fixed point of the $\Z_2$ twist,
and therefore a unique twist field.}
of $(x^0,x^3)$ by $\sigma$; it combines with the left-moving $U(1)$
ghosts into a conformal dimension zero
twist field $\Sigma=\sigma e^{\tilde\phi/2}$,
where $\tilde b\tilde c=\d\tilde\phi$ is the number current of
the $U(1)$ ghosts.
Also denote by $S^{i}_{\beta}$, $i=2,3$ the spin fields of the 
fermions $\lambda^a_i$, respectively.

The spectrum of the theory includes the eight scalars 
$V^a$ \scalrs, $8$ left-moving fermions $\bar\chi_{\bar\alpha}$ \frms,
and in addition $16$ right-moving chiral bosons (described
as R-R fields living at the fixed point of the $\Z_2$ twist)
\eqn\rwm{
  W^i_\beta=e^{-{1\over2}\phi+{i\over2}H}S^i_\beta\Sigma\;
	e^{-\frac12{\bar\phi_+} -\frac12{\bar\phi_-}
	+ \frac i2\bar H_2 -\frac i2\bar H_1}
	e^{ik\cdot x}\quad,\qquad i=2,3\ .
}
It is now important to include a factor of $\tilde c$ in the
vertex operators $V^a$, $\bar\chi_{\bar \alpha}$, which we have
suppressed until now.
Note that the $W^i_\beta$ have non-local operator product expansions 
with the supercurrents $Q_{\bar\alpha}$ \schg\ and with the right-moving 
fermions $\chi_\alpha$ \frms; hence the latter are eliminated 
from the spectrum. 
Note also that $W^i_\beta$ are inert under the unbroken (8,0)
supersymmetry, in agreement with their chirality.

Alternatively, one might try to replace these 
chiral bosons by 32 right-moving fermions.  To do this, compactify
$x^3$ on a circle whose radius is fixed by consistency; there
are then two fixed points of the $\Z_2$ twist, with corresponding
twist fields $\Sigma^{(s)}$, $s=1,2$.  We can then write 
vertex operators
\eqn\rwma{
  F^{i,(s)}_\beta=e^{-{1\over2}\phi+{i\over2}H}S^i_\beta\Sigma^{(s)}\;
	e^{-\bar\phi_+ - \bar\phi_-}
	e^{ik\cdot x}\quad,\qquad i=2,3\quad,\qquad s=1,2\ .
}
This representation is more in the spirit of \horwit, as half
of the right-moving current algebra arises from each fixed point.
It is not clear whether one of these two representations is
preferred.

The massless fields $V^a$, $\bar\chi_{\bar\alpha}$ and 
$W^i_\beta/F^{i,(s)}_\beta$ are free, as 
before, and apparently coincide with the field content of the heterotic 
string on the world sheet in the static gauge Green-Schwarz formalism.
Quantization of the theory should reveal that the chiral scalars 
$W^i_\beta$
live on either the $E_8\times E_8$ or ${\rm Spin}(32)/Z_2$ torus, due to
cancellation of gravitational anomalies in the target space of the $(2,1)$
string, but as explained previously, to see the torus more work is needed.
If the $\{V^a\}$ are compact, the 
target of the target string theory will live 
at some point on the Narain moduli space $SO(8,24)/SO(8)\times SO(24)$.


\newsec{Target Space Membranes}

In the previous section we saw that the $1+1$ dimensional 
target space version of the $(2,1)$ string gives rise
to critical (super-) strings. In this section we will demonstrate
that the $1+2$ dimensional version gives (super-) membranes.
Unlike strings, it is not known how to quantize membranes, and one
can imagine that the $(2,1)$ string will teach us.
We start with the construction of the supermembrane.

\subsec{The eleven-dimensional supermembrane.}

In section {\it 3.2} we have constructed a type II world sheet
starting from a $(2,1)$ string with the internal left-moving 
superfields $(y^a, \lambda^a)$, $a=1,\cdots,8$, and an $N=1$ 
superconformal current given by \spc. The scalars $y^a$ live on 
the $E_8$ torus, and the left-moving gauged $U(1)$ current was
chosen to be $J=\partial x^0+\partial x^3$. This produced a theory
on the $1+1$ dimensional target space 
parametrized by $(x^1, x^2)$, which we identified
with the type II world sheet. As pointed out in \ovthree,
there is a second possible choice: gauging a left-moving $U(1)$ 
null current $J$ whose timelike piece 
lies in $\IR^{2,2}$, and whose spatial
part lies in the internal space (in this case the $E_8$ torus),
such as:
\eqn\jmemb{J=\partial x^0+\partial y^1\qquad,\qquad
  J\ket{\rm phys}=0\ .}
If $x^0$ is initially non-compact, \jmemb\ effectively
compactifies it, identifying the momentum in the $x^0$
direction with the (quantized) momentum in the $y^1$ direction.
Therefore, the theory lives in $1+2$ non-compact 
dimensions, $(x^1, x^2, x^3)$. 

The target space theory is supersymmetric. To construct the supercharges
it is convenient to bosonize the left-moving fermions in a different
way than before:
\eqn\newbos{\eqalign{
\psi^1\pm\psi^2=&e^{\pm i H_1}\cr
\psi^3\pm i\lambda^2=&e^{\pm i H_2}\cr}}
and $(\lambda^3,\cdots,\lambda^8)$ bosonized in the standard
way. Note that the fact that the $U(1)$ current \jmemb\ is gauged,
$J|{\rm phys}\rangle=0$, is related by the superconformal algebra
\spc\ to the fermionic constraint
\eqn\fermrel{(\psi^0+\lambda^1)|{\rm phys}\rangle=0\ .}
Thus, $\psi^0$, $\lambda^1$ decouple, and can (and will) be ignored,
together with the spin $1/2$ bosonic ghosts that accompany \fermrel.

The spacetime supersymmetry
generators are constructed in the standard
way out of the $SO(4)$ spinors $S_A, S_{\bar A}=e^{{i\over2}
(\pm H_1\pm H_2)}$ ($A\in 2$, $\bar A\in \bar 2$ of $SO(4)$), and the
$SO(6)$ spinors $S_\alpha$, $S_{\bar\alpha}$ ($\alpha\in 4$, 
$\bar\alpha\in\bar4$) constructed out of $\lambda^3,\cdots,
\lambda^8$. 
\eqn\ssmem{
\eqalign{
  Q_{A\alpha}=&\oint e^{-\half{\phi}}S_A S_\alpha\cr
  Q_{\bar A\bar\alpha}=&\oint 
	e^{-\half{\phi}}S_{\bar A} S_{\bar\alpha}\ .\cr 
}}
The supercharges \ssmem\ form as before a $16$ of $SO(10)$, and decompose
as $(4,2)\oplus(\bar 4,\bar 2)$ of $SO(6)\times SO(4)$.
The superalgebra is:
\eqn\supalg{\{Q_{A\alpha}, Q_{\bar B \bar\beta}\}=\delta_{\alpha\bar\beta}
\gamma_{A\bar B}^\mu P_\mu}
where $P_\mu$ is the momentum in the $(x^1, x^2, x^3, y^2)$ 
directions, and $\gamma^\mu$ are four-dimensional $\gamma$ matrices
(essentially $\sigma$ matrices). The last component of the momentum 
is discrete and actually vanishes for the physical on-shell states, since
$\vec p_y$ belongs to an even self-dual lattice (the $E_8$ root lattice),
and states with $\vec p_y\not=0$ cannot satisfy $L_0=\bar L_0$.

There are two kinds of massless bosonic states: 
\eqn\vbos{
 V^a=e^{-\phi}\lambda^a e^{-\bar\phi_+-\bar\phi_-}
	e^{ik\cdot x};\;a=2,\cdots,8.}
and 
\eqn\vecxi{A=e^{-\phi}\xi\cdot\psi e^{-\bar\phi_+-\bar\phi_-}
	e^{ik\cdot x}}
The momentum $k$ and polarization vector $\xi$ are three dimensional;
the states $V^a$ \vbos\ are  seven scalars, while $A$ \vecxi\ is
a three dimensional gauge field. In $1+2$ dimensions the gauge field has
one propagating degree of freedom; it is dual to a scalar via the usual
$\epsilon^{\mu\nu\rho}F_{\nu\rho}=\partial^\mu\phi$.

There are also 8 fermions related to $V^a$, $A$ by supersymmetry
\ssmem.  
The target space theory is a theory of $2+1$ dimensional supergravity
coupled to eight matter degrees of freedom.
This system can be thought of as the world-volume description
of an eleven-dimensional supermembrane in a physical gauge.
The spectrum \vbos, \vecxi\ (and supersymmetry structure
\supalg) is identical to the (massless) spectrum of modes
living on a Dirichlet two-brane in type IIA 
string theory in ten dimensions
\ref\polch{J. Polchinski, hep-th/9510017, \prl{75}{1995}{4724}.}.  
There, the analogs of $V^a$ are the seven collective modes
for transverse fluctuations of the two-brane, while the analog
of $A$ is a gauge field on the world-volume of the two-brane.
The interactions in our case are somewhat different:
While for D-branes the analogs of $V^a$ are neutral with respect to
the gauge field, here there is a three-point coupling
between $A$ and two $V^a$, as well as other couplings of the bosons
\vbos, \vecxi\ to the fermions.  Of course, in addition
the Dirichlet two-brane has an infinite tower of massive excitations, 
and interacts in a highly nontrivial way with the modes living in the
bulk of spacetime, whereas we only find massless modes
and no states living away from the `brane'.
Furthermore, while in the construction of \polch\
the type IIA string is treated as fundamental and the 
Dirichlet two-brane is a soliton, we naturally regard the two-brane
as fundamental and the IIA string is a limit of it
under double dimensional reduction.
The relation between the Dirichlet two-brane of the IIA theory
and the membrane of eleven-dimensional supergravity has been
discussed in 
\nref\townsend{P. Townsend, hep-th/9512062.}%
\nref\schmidhuber{C. Schmidhuber, hep-th/9601003.}%
\refs{\townsend,\schmidhuber}.

We will leave a detailed exploration of the 1+2 dimensional
dynamics of our theory to future work.  It is likely that the
theory describes a membrane in a 10+1 dimensional space
in a physical gauge, with $8=11-3$ physical bosonic, and 8 physical
fermionic modes.
Note that while the theory lives in 1+2 non-compact directions,
it should really be thought of as living in 2+2 dimensions
with one of the time coordinates compact.  Due to 
\jmemb, the on-shell physical states \vbos, \vecxi\ carry
no momentum in the $x^0$ direction, but if we use the \twoone\
string to quantize the theory, states with $k_0\ne0$
would circulate in loops.  Perhaps reinterpreting
the standard 1+2 dimensional membrane as a 2+2 dimensional theory
will aid in the development of a quantum world-volume theory.

\subsec{The bosonic membrane.}

To see the importance of the 2+2 dimensional interpretation
of our theories, it is useful to construct explicitly
the `bosonic membrane' -- the 1+2 dimensional analog
of the theory considered in section {\it 3.1}.

Thus we write our \twoone\ string theory in terms of 24 fermions
$\lambda^a$, with the superconformal current \leftNeqone.
We need to isolate a $U(1)$ supercurrent to gauge in this sector.
To do that, choose
\eqn\gint{
  G_{int}=\lambda^1\lambda^2\lambda^3 + \coeff16 f_{abc}
	\lambda^a\lambda^b\lambda^c\ ,
}
with $a,b,c=4,...,24$ and $f_{abc}$ the structure constants
of a 21-dimensional Lie algebra. The gauged $U(1)$ supercurrent
is then taken as
\eqn\jpsi{\eqalign{
  J(z)=&\d x^0+\lambda^2\lambda^3\quad;\qquad J\ket{\rm phys}=0\cr
  \psi(z)=&\psi^0+\lambda^1\quad;\qquad \psi\ket{\rm phys}=0\ .\cr
}}
Equation \jpsi\ relates $k_0$ to the $\lam^2$, $\lam^3$
fermion number.  The physical states consist of 21 massless scalars
\eqn\twone{
  V^a=e^{-\phi}\lam^a e^{-\phibar_+-\phibar_-}e^{ik\cdot x}
	\quad,\qquad a=4,...,24; 
}
in addition one finds
\eqn\vpm{\eqalign{
  T^+=&e^{-\phi}(\lam^2+i\lam^3)
	 e^{-\phibar_+-\phibar_-}e^{i x^0}e^{ik\cdot x}\cr
  T^-=&e^{-\phi}(\lam^2-i\lam^3) 
	e^{-\phibar_+-\phibar_-}e^{-i x^0}e^{ik\cdot x}\ ;\cr
}}
these states are scalars that are massless in four dimensions, but
in 1+2 dimensions they are tachyonic with $m^2=-1$:
\eqn\tach{E^2=(k_1)^2=k_2^{\ 2}+k_3^{\ 2}-1}
because they carry quantized momentum in the timelike $x^0$ direction.
Finally, we have a three-dimensional gauge field
\eqn\gf{
  A=e^{-\phi}\xi\cdot\psi e^{-\phibar_+-\phibar_-}e^{ik\cdot x}
}
with $\xi\cdot k=0$, $\xi\approx\xi+\epsilon k$ as before.

The fact that $T^\pm$ are tachyons in 1+2 dimensions
does not lead to infrared divergences in the one-loop amplitude of the
\twoone\ string, as one would ordinarily expect in
a field theory with tachyons.
One simply has to revert to the 2+2 dimensional
view of the theory.  This seems to indicate that we should
think of the theory not as a membrane in 26+1 dimensions,
but rather as a theory of a membrane with two world-volume time
coordinates, one of which is compact, living in 26+2
dimensions.  Presumably the same holds for the supermembrane,
which should then be thought of as having a worldsheet
of signature 2+2 and living in 10+2 dimensional spacetime.


\newsec{General Backgrounds}

So far we have concentrated on the string in a flat background;
however, the topological perturbation theory of the target
string or membrane requires us to consider the N=\twoone\
theory on a curved target space.  To that end, let us review the
beta function equations of \twoone\ sigma models
\ref\hullNeqtwo{C.M. Hull,
\pl{178}{1986}{357}.},\ref\bonneau{G. Bonneau and G. Valent,
\cqg {11}{1994}{1133}; hep-th/9401003.}.

The \twozero\ superspace description of N=\twoone\ heterotic
strings involves real scalar chiral superfields $\XX^\mu$,
$\mu=0,1,2,3$ with signature $(-,-,+,+)$.  Pairing these into
light-front coordinates
\eqn\lightfront{\eqalign{
  X^1=\XX^0+\XX^3\qquad&\qquad X^{\bar 1}=\XX^0-\XX^3\cr
  X^2=\XX^1+\XX^2\qquad&\qquad X^{\bar 2}=\XX^1-\XX^2\ ,\cr
}}
the needed \twozero\ superfields are
\eqn\Xfield{
  X^i(z;\zbar,\tbar_\pm)=x^i
        +\tbar_+\psibar^i +\cdots\quad,\qquad
  X^\ibar(z;\zbar,\tbar_\pm)=x^\ibar
        +\tbar_-\psibar^\ibar +\cdots\ ,
}
satisfying $\Dbar_-X^i=\Dbar_+X^\ibar=0$.
In addition, there are 28 left-moving fermionic superfields
$\Lambda^a(z;\zbar,\tbar_\pm)= \lambda^a +\cdots$
satisfying $\Dbar_-\Lambda^a=W^a$ for some superfield with
$\Dbar_-W^a=0$.
The action is
\eqn\action{
  S=-\hf\int d^2z\;d^2\tbar\biggl[
        i\Bigl(K_i(X,\Xbar)\d X^i - K_\ibar(X,\Xbar)\d X^\ibar\Bigr)
        +V_{a^* b}(X,\Xbar)(\Lambda^a)^*\Lambda^b
        \biggr]\ .
}
Consistency requires 
(a) that the target space parametrized
by $X$, $\Xbar$ have signature 4+0 or 2+2 (we will concentrate
on the latter); 
(b) that there exist
a holomorphic supercurrent
$G=e_{\mu a}(\XX)\Lambda^a\d \XX^\mu +
f_{abc}(\XX)\Lambda^a\Lambda^b\Lambda^c$; and
(c) that we gauge an anomaly-free, left-moving
current $\JJ=\VV_a(\XX) \Lambda^a$ and its holomorphic superpartner.

In the free theory (flat spacetime), the background is
\eqn\flatspace{
        K_i=\delta_{i,\ibar}X^\ibar\quad,\qquad
        K_\ibar=\delta_{\ibar,i}X^i\quad,\qquad
        V_{ab}={\rm const.}\ .
}
The N=1 left-moving supercurrent is obtained by splitting
off four of the twenty-eight $\Lambda$'s, call them $\Psi^i$,
$\Psi^\ibar$, and pairing them with the $X^i$, $X^\ibar$.
Then $e_{\mu a}=\delta_{\mu a}$, and a solution for $f_{abc}$
may be determined by putting the remaining $\Lambda^a$
in the adjoint of some semisimple group of
dimension 24 \pierce\
(for which the $f_{abc}$ are the structure constants).
Note that \flatspace\ looks like the embedding of a (2+2) -- brane
into a 26+2 dimensional spacetime.

The semi-Kahler potential $K_\mu$ of \action\ determines the metric
and torsion in holomorphic coordinates:
\eqn\geom{
  g_{i\jbar}+b_{i\jbar}=K_{i,\jbar}\qquad
  g_{ij}=b_{ij}=0\ .
}
It proves convenient to use the connection with torsion 
$\Gamma^\mu_{\ \nu\lambda} = \gamma^\mu_{\ \nu\lambda} - \hf 
T^\mu_{\ \nu\lambda}$ ($\gamma$ is the Christoffel connection).
The torsion is determined as
\eqn\torsion{
  T_{ij\kbar}=\d_\kbar(K_{i,j}-K_{j,i})\qquad\qquad T_{ijk}=0\ ,
}
and the one-loop beta function equations 
$R^{(-)}_{\mu\nu}=D_\mu D_\nu\phi$ boil down to
\eqn\betafns{
  \eqalign{V_i=T^k_{ki}=&\hf \d_i\phi^*\cr
  	V_\ibar=T^\kbar_{\kbar\ibar}=&\hf \d_\ibar\phi\cr
	\exp[\phi+\phi^*]=&\det[g]\cr
	g^{i\jbar}D_i\d_\jbar\phi=&g^{\ibar j}D_\ibar\d_j\phi^*=0\ .\cr
}}
Here $\phi+\phi^*$ is the dilaton of the \twoone\ string,
and $\phi-\phi^*$ is the theta parameter of its abelian
world sheet gauge theory.  Note that the equations are invariant
under the gauge transformation
\eqn\gaugesym{
  \delta K_i=\d_i\chi\quad,\qquad \delta K_\ibar= -\d_\ibar\chi\ ;
}
hence, counting degrees of freedom in the gravity sector, we
should have four -- two from the dilaton and theta fields,
and two from $K_\mu$ (four minus one gauge invariance and one
gauge condition).  The worldsheet fields $\Lambda$ give rise to
self-dual Yang-Mills in the 2+2 target space, parametrized by
a single adjoint scalar field (\eg\ 24 scalars for the
`bosonic string' analog).  At one loop, anomaly cancellation
requires $dT=\tr[R^2-F^2]$.
The Yang-Mills field strength $F$ enters the beta function
equations first at two loops; 
one finds curvature-squared corrections
to \betafns\ \hullNeqtwo:
\eqn\twoloop{
  R^{{\sst (-)}}_{\;\mu\nu}\longrightarrow R^{{\sst (-)}}_{\;\mu\nu}
	+\frac{\alpha'}4\Bigl[{R^{{\sst (+)}\;\rho\sigma\tau}}_\mu
	R^{\sst (+)}_{\;\rho\sigma\tau\nu} - 
	F_{\mu\rho}F_\nu^{\ \rho}\Bigr]\ ,
}
where the superscripts denote the curvature of the connection
$\Gamma^{\sst (\pm)}=\gamma\pm T$.
If we want to regard the field space 
of this theory as the target space of a 2+2 world volume,
our off-shell counting of target space dimension agrees with 
the expectation generated by the analysis of sections three and
four -- $dim(G)$ (\eg\ twenty-four) transverse coordinates together
with 2+2 longitudinal coordinates from the gravity sector.
Note that $dT-\tr[R^2-F^2]=0$ looks a bit like a
Virasoro constraint; indeed, it comes from a reparametrization
anomaly in the sigma-model field space.

We have not analyzed the conditions for a left-moving
null current algebra and supersymmetry.  In the case of 
reduction to 1+1 dimensions the appearance of a current algebra
should amount to the requirement 
that the 2+2 geometry admit a covariantly constant
null Killing vector.
Target space supersymmetry requires the \twoone\
sigma model carry an N=(2,2) global supersymmetry, which
further constrains the semi-Kahler potential; the torsion
connection $\Gamma$ is embedded in the gauge group.
The details of the sigma model geometry
may help uncover the 10+2 supergravity theory hinted at
by our work.


\newsec{Comments and Speculations}

Our results open a number of intriguing avenues of investigation.
Among the potential applications of our work are:

\item{i)} As an indication of substructure to world-volumes.
This was one of the main motivations of \green.  Here,
one finds a sequence: World sheet $\hookrightarrow$
self-dual four-space $\hookrightarrow$ spacetime.
Since different string theories are related on the first level
by orbifold and deformation of the null current, and on the
last level by  strong/weak coupling duality, one might wonder
whether there is another link: $(2,1)$ world sheet $\leftrightarrow$
spacetime.

\item{ii)} As an indication of how to construct the quantum theory
of membranes directly as a quantum field theory
in 2+2 dimensions.  One imagines using
the (2,1) string as a sort of scaffolding to construct such a theory,
to be removed upon completion.

\item{iii)} As a novel construction of soliton strings/membranes
describing critical string/M-theory compactified down to two or three
dimensions.

\noindent
Let us comment on these and a few related topics. 

\subsec{The N=(2,0) string.}

The $(2,0)$ string is a theory obtained by combining
a right moving $N=2$ string with a left moving bosonic
string (see e.g. \ovthree). The simplest vacua
correspond to compactifications of the left movers
on one of the $24$ Niemeier tori. In his original
paper, Green pointed out that the theory obtained by
compactifying on the Leech torus is a critical bosonic
string. It would be interesting to understand these theories
in general. Like the $(2,1)$ strings discussed above,
they only have massless excitations, and when the target
space is $1+1$ dimensional, these are free.

There are two differences:
\item{a)} Even in the simplest constructions mentioned
above, the $(2,0)$ string generically gives rise to 
{\it non-critical} string theories in target space.
Indeed, the spectrum of the theory on a particular Niemeier
lattice includes the $24$ oscillator modes $\partial y^a$,
and in addition momentum states $\exp[i\vec p\cdot \vec y]$ with
$p^2=2$. These massless states form the adjoint representation
of a rank $24$ group $G$ which depends on the lattice (for
the Leech lattice the group is $U(1)^{24}$ but in general it is more
complicated).

\item{b)} The decoupling of ``momentum'' and ``winding'' strings,
described for the $(2,1)$ case in section 3 (see \resz) does not occur
here. The analog of the expression for $Z_1$ \zone\ in this case
is:
\eqn\znzero{
  Z_1(T,U)={1\over2}\int_\FF {d^2\tau\over\tau_2^2 }
         \left(J(\tau)+C_0\right)
	\sum_{(p_L, p_R)}q^{{1\over2}p_L^2}
	\bar q^{{1\over2}p_R^2}
}
where $J(\tau)=1/q+\cdots$ is the unique meromorphic
modular invariant function with the specified behavior
as $q\to0$ and no constant term in the expansion in powers of
$q$. $C_0$ is the number of massless states, equal to the dimension
of the group $G$ corresponding to the particular Niemeier
lattice.

\noindent
This integral has been computed in 
\ref\jfunction{G. Lopes Cardoso, D. L\"ust, and T. Mohaupt,
\np{450}{1995}{115}; hep-th/9412209.}, with the result:
\eqn\hmres{Z_1(T,U)=-C_0\log\left(\sqrt{T_2}|\eta(T)|^2
	\sqrt{U_2}|\eta(U)|^2\right)-\log|J(T)-J(U)|^2.}
The first term on the right hand side is just what one would
expect for a pair of non-critical strings living on the
$T$ and $U$ tori as for  the $(2,1)$ string discussed  in section 3.
The second term on the r.h.s. means (comparing 
\hmres\ to \zzone) that the decoupling between
$T$ and $U$ that took place there does not occur for the $(2,0)$
string, probably because of the lack of an $N=2$ superconformal 
structure in the left moving sector. It would be interesting to
understand these theories better.

\subsec{Strong/weak coupling duality.}

The S-duality group of the \twoone\
string is the T-duality group of its target string theory;
thus \twoone\ strings may serve as a useful laboratory for
probing the structure of S-duality in string theory, given that
we understand T-duality rather well.
We also have yet another context in which to investigate
target space topology change, given that eventually one will
have to sum over the topology of the target space.   This is
certainly true in the 1+1 dimensional reduction, and likely
to be true in the 2+1 reduction as well. 
There is also the possibility, given that we are on the road
to an explicit description of quantum membranes, that we will
understand the rules for constructing orbifolds of M-theory
(a hint of which we saw in section 3.3),
the T-duality structure of M-theory, and so on.

Hull \hull\ has pointed out that a twelve-dimensional origin
of string theory would explain the $SL(2,\Z)$ S-duality of
type IIB strings in ten dimensions as the T-duality symmetry
of a reduction from twelve to ten dimensions on
a two-torus.  One would like to see how that arises in our
framework.

Different choices of GSO projection of the \twoone\ string give
rise to different critical (super-) string theories.  Yet we
know another way that these theories are connected: Strong-weak
coupling duality.  Could it be that the S-duality group
of the various string and M-theories is related to
the discrete symmetry group of \twoone\ strings?  This group includes
the group of GSO projections.  One would
also need to include the discrete choice of null current 
orientation in such a construction.
This exciting possibility
would close the chain of embeddings of parameter space into
field space that makes spacetime from 2+2 world volume from
\twoone\ string worldsheet.

\subsec{Analogies with soliton strings.}

There is another possible interpretation of our construction
based on its similarity to that of solitonic
strings and other $p$-branes in higher-dimensional string theory \polch,
\ref\jh{J. Harvey and A. Strominger, hep-th/9504047, \np{449}{1995}{535};
A. Sen, hep-th/9504027, \np{450}{1995}{103}; D. Kutasov, hep-th/9512145.}. 
Indeed, the target geometry we described was
that of a $p$-brane world sheet stretched across all
of $p$+1 dimensional spacetime, for $p=1,2$.  The solution breaks
half of the supersymmetries, which arise in a Green-Schwarz
type formalism.  The tension of the target brane is related to the
string coupling of the \twoone\ string, and when the target space
is two dimensional, the coupling of the target string goes
to infinity as the coupling of the $(2,1)$ string goes to zero.

In higher dimensional string theories, soliton string
constructions often point to strong-weak coupling dual pairs.
One may ask whether it is possible that this is also the case here.
Specifically, could it be that a strongly coupled ten dimensional
heterotic or type II string compactified on an $8$ torus
is equivalent to a weakly coupled two dimensional $(2,1)$ string?
At first sight this seems ludicrous: a heterotic or type II string
theory has, even in a two dimensional spacetime, an infinite
tower of states (with exponential density of states at large mass), 
while the $(2,1)$ string describes a small collection of massless
two dimensional field theoretic degrees of freedom with no
massive states at all. 

One may ask, what happened to the 
exponential density of BPS states of the critical string?
By definition, they are all charged under some U(1)
gauge group, which confines in two dimensions;
there are no BPS states in the physical asymptotic spectrum!
What remains are the `moduli fields' and gravity.
Gravity merely dresses the
renormalization group flows of various operators.
The moduli fields 
parametrize a nonlinear sigma model; 
they are disorderd in two dimensions; 
the theory does not
have a manifold of ground states. The moduli sigma model
is infrared free (at weak coupling) -- the theory flows towards 
$8\times 8$ or $8\times 24$ free fields coupled to dilaton
gravity, \ie\ a noncritical string. This seems in disagreement
with the massless spectrum of the $(2,1)$ string, which 
contains $8$ right moving and either $8$ or $24$ left
moving scalars.  However, we saw in
section 3 that when the \twoone\ string is weakly coupled,
the target theory is very strongly coupled; it is not completely
clear what the sigma model dynamics is in this situation.  
In fact, examples
are known in N=1 supersymmetric Yang-Mills in four dimensions
\ref\kutasov{D. Kutasov, A. Schwimmer and N. Seiberg, 
hep-th/9510222.}
whereby a naively irrelevant flow at weak coupling is
nevertheless relevant at strong coupling and drives the
theory to a nontrivial fixed point.  Could that be happening
here?  It seems suggestive that the counting of moduli fields
of toroidally compactified superstrings
agrees with that of mesons made from the target
fields of the \twoone\ string.

A similar discussion would lead one to conclude that the
bosonic string that is well known to be perturbatively 
unstable at weak coupling, is in fact stable at strong coupling
(at least when compactified to $1+1$ spacetime dimensions), and is
equivalent to a version of the $(2,1)$ string,
described in section 3.

If the above discussion is valid, it implies that strongly 
coupled $2d$ strings have vastly fewer degrees of freedom than
one would deduce perturbatively, an idea that has been suggested
in various contexts in the past.

\subsec{New principles.}

The reduction of four-dimensional self-duality to yield
both string and membrane world-volume theories suggests that one
might elevate self-duality to a principle underlying unification,
superseding the principle of two-dimensional conformal invariance 
which has been
so fruitful in developing our understanding of strings.
To this end, we need to generalize our treatment above, which 
took place in a physical gauge where spacetime Lorentz invariance 
and half the supersymmetries are nonlinearly realized.
Similar issues have arisen in the effort to find an acceptable
covariant quantization of the Green-Schwarz string formalism;
there, the use of twistor methods \ref\twistor{
F. Delduc, A. Galperin, P. Howe, and E. Sokatchev,
\physrev{47}{1993}{578}; hep-th/9207050.
} 
has helped elucidate the 
geometrical structure.  Roughly, the twistor variables 
covariantize the breaking of $SO(9,1)$ Lorentz symmetry 
down to the little
group $SO(1,1)\times SO(8)$, and at the same time 
bring in fascinating connections to Hopf fibrations, Jordan algebras,
octonions, lightlike integrability, etc.  
A natural starting point might be to generalize
these methods to the conformal group $SO(10,2)$, which also
has a twistorial construction.  The null gauging that appears
in the N=\twoone\ string might have an interpretation in terms of
group contractions of this larger (super-) symmetry.
Note, in particular, that central charges arise in this way 
\ref\lukierski{J. Lukierski and L. Rytel,
J. Phys. {\bf A15} (1982) L215.}.

We do not necessarily mean to suggest that the only way to obtain a
quantum theory of membranes is to construct a string theory
whose target space is the membrane world-volume.  Rather, one
might interpret the \twoone\ heterotic string as a guide toward
the construction of an appropriate world-volume quantum field theory.
It is not out of the question that one could quantize directly
the reduced version of self-dual gravity coupled to self-dual
matter that appears as the effective target dynamics of the
\twoone\ string.  First of all, the fluctuations of the geometry
are restricted to be self-dual even off-shell; second, the null
gauging of the \twoone\ string leads to further restriction
on the target fluctuations.  These may be enough to render the
theory renormalizable.  Note that the usual string effective Lagrangian
$e^{-2\phi}\sqrt{g}[R+\cdots]$ leads to improved power counting
if we use $g_{i\jbar}=\d_{(i}K_{\jbar)}$ and treat $K$ as the
fundamental dynamical field.  Constraints on matter content (\ie\ 10+2
or 26+2 dimensions of spacetime) should arise from quantum anomalies in
self-dual gravity Ward identities in 2+2 dimensions.  

If there is secretly a second timelike coordinate
that only appears in off-shell dynamics, what are its
consequences? Could there be novel mechanisms for solving
some of the longstanding problems of unification?  
Supersymmetry breaking, the
cosmological constant, and other hierarchy problems come to
mind; as well as the horizon problem in cosmology and the
Hawking paradox.  
After all, the causal structure of spacetime is subtly different
in this theory; and the second, off-shell time direction could allow
communication between vastly different
energy and time scales of the principal time coordinate
via fluctuations travelling along the second time.
Another fascinating feature of the \twoone\ string
is the joint appearance of two apparently decoupled string theories
living in mirror target spaces; can they be coupled (perturbatively
or nonperturbatively), as they are in \twozero\ strings?  
What would this mean?

\subsec{Related ideas.}

The constructions in this paper touch on a number of ideas
about underlying structure in string theory
that have been proposed over the years:

\item{a)} The complexification of the world sheet and spacetime
\ref\wittencomplex{E. Witten, \prl{61}{1988}{670}.}.  These
have been considered before, in studies of high-energy/temperature
behavior of strings.

\item{b)} Worldsheet -- target space duality.  Giveon et.al.
\ref\giv{A. Giveon, N. Malkin and E. Rabinovici, \pl{ 220} {1989} {551}.}
noticed that a string compactified on a target torus exhibits
a similarity between the dependence on the Narain moduli of
the target space and the world sheet period matrix.

\item{c)} The possibility of a hidden relation between
bosonic and fermionic strings.  Freund \ref\freund{
P. G. O. Freund, 
\pl {151}{1985}{387}.  See also
A. Casher, F. Englert, H. Nicolai, and A. Taormina,
\pl{162}{1985}{121}; 
F. Englert, H. Nicolai, and A. Schellekens,
\np{274}{1986}{315}. }
was the first to suggest
that one might be able to obtain the fermionic string from
the bosonic string.

\item{d)} That the `number of
degrees of freedom' in string theory (whatever that means)
is vastly reduced at short distances 
\ref\susskind{I. Klebanov and L. Susskind, \np{309}{1988}{175}.}, 
or high temperatures \ref\atickwitten{J. Atick
and E. Witten, \np{310}{1988}{291}.}
has been suggested by
several independent approaches; there are also indications
of two-dimensional structure in the asymptotic spectrum
\ref\kutseib{D. Kutasov and N. Seiberg, \np{358}{1991}{600}.}.
These ideas resonate strongly with the interpretation of
the \twoone\ string as a strong-coupling dual of critical
string theories, given its two-dimensional
field-theoretic density of states.

\item{e)}
We have seen in sections 3 and 4 (see also \ovthree)
that the \twoone\ string secretly does live in \tpt\ dimensions,
in some instances with one of the time directions compactified.
This leads to the possible emergence of hyperbolic algebras
as a hidden symmetry of string/membrane theory; for instance, with
one space and one time direction of the \tpt\ spacetime compactified,
the right-moving dynamics involves $\Gamma_{1,1}$, while that
of the left movers takes place in $\Gamma_{9,1}$.  Structures of this
sort have been considered in 
\nref\moore{G. Moore, hep-th/9305139.}%
\nref\harveymoore{J. Harvey and G. Moore, hep-th/9510182.}%
\refs{\moore,\harveymoore}.
Note that there is a unique null vector in such lattices;
this may lead to a common treatment of strings and membranes
within our framework.

\item{f)} 
More than 10 or 11 dimensions.  Attempts have been made before
to go beyond the barrier of eleven as the maximal dimension
of supergravity \ref\beyond{L. Castellani, P. Fre, F. Giani,
K. Pilch, and P. van Nieuwenhuizen, \physrev{26}{1982}{1481};
M. Wang, {\it Supergravity in twelve dimensions}, 
in the Proceedings of the Marcel Grossmann Meeting, Rome 1985,
p. 1459; P. G. O. Freund, {\it Introduction to Supersymmetry},
Cambridge Press, 1986, ch. 26 (!);
M. Blencowe and M. Duff, \np{310}{1988}{387}.}.
Hull recently presented indirect evidence for the existence of such
a theory on the basis of lower-dimensional vacua of string/M-theory
\hull; further evidence was presented in 
\nref\vafa{C. Vafa, hep-th/9602022; D. Morrison and C. Vafa,
hep-th/9602114.}%
\nref\tseyt{A. A. Tseytlin, hep-th/9602064.}%
\refs{\vafa,\tseyt}.


\vskip 1in
{\sl Acknowledgements}:
It is a pleasure to thank 
M. Douglas,
P.G.O. Freund, 
J. Harvey,
and M. O'Loughlin
for discussions.

\listrefs
\bye